\documentclass[12pt]{article}
\usepackage{amsmath, amssymb, mathrsfs, mathtools, tensor, braket}
\usepackage{xcolor, graphicx, subfigure}

\numberwithin{equation}{section}

\usepackage{enumitem, float}

\usepackage{caption}
        \DeclareCaptionFormat{myformat}{#3}
\evensidemargin \oddsidemargin

\begin{document}
        
        \vspace{5mm}
        \begin{center}
                {{\Large \bf Inhomogeneous Abelian Chern-Simons Higgs Model\\with New Inhomogeneous BPS Vacuum and Solitons}}
                \\[17mm]
                Yoonbai Kim, ~~O-Kab Kwon, ~~Hanwool Song \\ [2mm]
                {\it Department of Physics, Sungkyunkwan University, Suwon 16419, Korea} \\ [-1mm]
                {\it yoonbai@skku.edu, ~okab@skku.edu, ~hanwoolsong0@gmail.com} \\ [3mm]
                Chanju Kim \\ [2mm]
                {\it Department of Physics, Ewha Womans University, Seoul 03760, Korea} \\ [-1mm]
                {\it cjkim@ewha.ac.kr}
        \end{center}
        \vspace{15mm}

        \begin{abstract}
                
        \end{abstract}
We study an inhomogeneous U(1) Chern-Simons Higgs model with a magnetic 
impurity in the BPS limit. The potential is sextic with both broken and 
unbroken phases, but its minimum varies spatially depending on the strength 
of the impurity. While the system lacks translation symmetry, it
admits a supersymmetric extension. Depending on the sign of the impurity term, 
it has either a BPS sector or an anti-BPS sector (but not both), which satisfies
the Bogomolny equations. The vacuum configuration of the broken phase is 
not simply determined by the the minimum of the potential since it is no longer 
constant, but it becomes a nontrivial function satisfying the Bogomolny 
equations. Thus, the energy and angular momentum densities of the vacuum 
locally have nonzero distributions, although the total energy and angular 
momentum remain zero. As in the homogeneous case, the theory supports various 
BPS soliton solutions, including topological and nontopological vortices and 
Q-balls. The vorticities as well as the U(1) charges are exclusively positive 
or negative. For a Gaussian type impurity as a specific example, we obtain 
rotationally symmetric numerical solutions and analyze their detailed 
properties. 
We also discuss the case of a delta-function impurity as the infinitely thin limit
of the Gaussian impurity which shows some nontrivial feature of BPS 
Chern-Simons Higgs theory.
        
\newpage
        
\section{Introduction}
Inhomogeneities in field theories can appear in various physical contexts, 
e.g. as external fields, defects/impurities or junctions of
heterogeneous systems in the experimental setup, or as theoretical probes 
to study some properties of the system.
Due to the lack of the translation symmetry in inhomogeneous theories, 
it is usually difficult even classically to analyze the system by analytic 
means.

This problem can be alleviated if the theory has a BPS sector that saturates
a Bogomolny bound, since Bogomolny equations are first-order differential 
equations~\cite{Bogomolny:1975de,Prasad:1975kr}. 
In the usual case without inhomogeneity, such theories allow for
supersymmetric extensions~\cite{DiVecchia:1977nxl}. 
From the supersymmetry algebra, it can be seen that the energy of BPS 
configurations is proportional to a central charge and the 
vanishing conditions of unbroken supercharges are identified with the Bogomolny 
equations~\cite{Witten:1978mh}. 

This is also true for inhomogeneous theories, although Poincare symmetry 
is explicitly broken. For example, Janus Yang-Mills theories with a 
position-dependent gauge coupling in four dimensions~\cite{Bak:2003jk}
are dual to dilatonic deformations of AdS$_5$ space and have supersymmetric 
extensions~\cite{Clark:2005te, DHoker:2007zhm, DHoker:2006qeo, Kim:2008dj, Kim:2009wv}. 
In three and four dimensions, mass-deformed ABJM and super Yang-Mills theories 
can have further inhomogeneous mass deformations, preserving some of the 
supersymmetries~\cite{Kim:2018qle, Kim:2019kns, Arav:2020obl, Kim:2020jrs}. 
There are also other supersymmetric theories with impurities in two and three 
dimensions~\cite{Hook:2013yda, Tong:2013iqa, Adam:2019yst}. 

Recently, we considered 1+1 dimensional classical supersymmetric 
inhomogeneous theories with a single scalar field as the simplest example 
of inhomogeneous theories, where the superpotential is allowed to have 
a spatial dependence that breaks translation invariance~\cite{Kwon:2021flc}.
Half of the supersymmetry in the homogeneous theory is preserved by adding a 
term which is a derivative of the superpotential with respect to the position. 
For certain types of inhomogeneities in theories such as sine-Gordon theory 
and $\phi^6$ theory, we have been able to obtain general solutions of the 
Bogomolny equation, independent of the detailed form of the spatial variation.
See also \cite{Ho:2022omx} for the relationship between supersymmetric
field theories on a curved background metric and supersymmetric inhomogeneous
field theories in 1+1 dimensions.

In this paper, we would like to study classically an inhomogeneous version 
of the self-dual U(1) Chern-Simons Higgs (ICSH) model in 2+1 dimensions.
The homogeneous theory~\cite{Hong:1990yh,Jackiw:1990aw} has been extensively 
studied in the context of anyons and fractional 
statistics~\cite{Wilczek:1981du,Arovas:1985yb},
and applied to condensed matter physics such as
fractional quantum Hall effect~\cite{Tsui:1982yy}
or anyon superconductivity~\cite{Chen:1989xs}.
In particular, with
a sextic potential having both broken and unbroken degenerate vacua, the 
system has a BPS sector and can be generalized to have $\mathcal{N}=2$ 
supersymmetry~\cite{Lee:1990it}. It has rich spectrum of soliton 
solutions~\cite{Jackiw:1990pr}: topological vortices,
nontopological solitons (Q-balls) and nontopological vortices (Q-vortices),
depending on the asymptotic behavior of the scalar field. They all have
nonzero spins, which is a characteristic of the Chern-Simons theory.

We can make the system inhomogeneous by deforming the vacuum expectation 
value $v$ of the scalar field in the broken phase to be a nontrivial function 
$v(\boldsymbol{x})$ of the position~\cite{Kwon:2021flc}. It explicitly breaks the translation 
symmetry but the BPS nature can be restored by adding a magnetic impurity term 
to the Lagrangian. Then it can also be extended to $\mathcal{N}=1$ 
supersymmetric theory, similar to the abelian Higgs model with impurities
\cite{Hook:2013yda, Tong:2013iqa}.
The inhomogeneous model considered here was first introduced
in~\cite{Han:2015tga} where the existence of topological multivortex solutions
was rigorously proved.

Inhomogeneous theories with a BPS sector
have in common that the sign of the inhomogeneous term 
$\Delta\mathscr{L}$ added to the Lagrangian can be either positive
or negative. This can be understood naturally in the supersymmetric extension
of the theory where half of the supersymmetries of the homogeneous theory 
are either broken or
unbroken, depending on the sign of $\Delta\mathscr{L}$. 
Therefore, in a certain inhomogeneous theory,
there can only be either a BPS or an anti-BPS sector, but not both. 
In the context of ICSH model, we have BPS vortices with either
positive or negative vorticities, but not both.
 
Since the naive ``vacuum expectation value'' $v(\boldsymbol{x})$ varies in 
space, the broken vacuum can no longer be a constant. 
In fact,
it is even not entirely clear whether there exists a broken vacuum solution 
with vanishing energy. We will show that the broken vacuum 
is given by a nontrivial configuration which is a solution of the Bogomolny
equations and has both vanishing energy and vanishing angular momentum. 
However, the energy and angular momentum densities turn out
to be locally nontrivial thanks to the Chern-Simons gauge field.
Since there is still the unbroken vacuum $\phi=0$ in the 
ICSH model, we have both broken and unbroken degenerate vacua in the theory
and hence the same type of solutions as in homogeneous case.

In this paper, we will mostly work with a rotationally symmetric Gaussian
impurity centered at the origin and numerically obtain various soliton
solutions. It is shown that most of the physical quantities of the solutions, 
such as energy and angular momentum are identical to those of the 
homogeneous theories, while the details are rather different due to the 
presence of the impurity.

We also consider the case of delta-function impurity as the infinitely thin 
limit of the Gaussian impurity. Due to the characteristic nature of the 
Chern-Simons field, the broken vacuum solution in this limit turns out to
be ill-defined at the impurity point. Away from the point, however, the 
delta-function impurity has no effect on the BPS solutions and the solutions
are exactly the same as those in homogeneous theory with no impurity.

The rest of the paper is organized as follows. In section 2, we introduce
the ICSH model with a magnetic impurity term and derive Bogomolny equations.
We also identify the supersymmetric Lagrangian and construct the supersymmetry 
algebra. In section 3, we study the vacuum configurations of ICSH model
in detail and obtain explicit solutions numerically for Gaussian impurities
and its limit to delta-function impurities.
In section 4, we obtain various topological as well as nontopological soliton
solutions. We conclude in section 5. 
There are two appendices. Appendix A discusses an off-shell
formulation of the supersymmetric ICSH model where electric impurities
are also briefly considered. In appendix B, we study another type of
regularized delta-function impurity and show that the value 
of the scalar field at the impurity point depends on the regularization.

\vspace{10mm}

{\bf Note added}: After submitting the paper, we became aware of the 
recent paper \cite{Bazeia:2024fgo} which has a partial overlap with our
work. It considers the same ICSH model with a Gaussian-type magnetic 
impurity and discusses some basic features of the vacuum in the broken phase
as well as topological soliton solutions. 

\section{BPS Limit of Inhomogeneous Chern-Simons Higgs Model}
        
In $(2+1)$ dimensions, abelian Chern-Simons Higgs (CSH) model is described by the Lagrangian density
\begin{equation}
                \mathscr{L}_{\rm CS} = \frac{\kappa}2 \epsilon^{\mu\nu\rho} A_{\mu} \partial_{\nu} A_{\rho} -\overline{D_{\mu} \phi} D^{\mu} \phi -  V(|\phi|)
                ,\label{092}
\end{equation}
where $\kappa>0$ and $\phi$ is a complex scalar field with covariant derivative
$ D_{\mu} \phi = ( \partial_{\mu} - i A_{\mu} ) \phi$ .
If the potential $V$ is given by
        \begin{align}
                V(|\phi|) = \frac{1}{\kappa^{2}} |\phi|^{2} (|\phi|^{2} - v^{2})^{2}
                ,\label{011}
        \end{align}
        which is sextic in $|\phi|$ with a coupling fixed by the Chern-Simons 
coefficient $\kappa$, the BPS bound is saturated \cite{Hong:1990yh,Jackiw:1990aw} and the theory admits a $\mathcal{N}=2$ supersymmetric extension \cite{Lee:1990it}. 
Note that the potential allows both symmetric and broken phases. In the former
with vanishing vacuum expectation value $\braket{|\phi|} = 0$, there is no propagating gauge mode, while two massive charged mesons propagate with mass
        \begin{align}
                m_{\bar{\phi}} = \frac{v^{2}}{\kappa} = m_{\phi}
                .\label{043}
        \end{align}
        In symmetry-broken phase of non-zero vacuum expectation value $\braket{|\phi|} = v$, the gauge boson of a single longitudinal degree and a neutral Higgs boson propagate with degenerate mass
        \begin{align}
                m_{A_{\mu}} = \frac{2v^{2}}{\kappa} = m_{\rm H}
                .\label{047}
        \end{align}
        
        In this paper we are mainly interested in the case that the parameter $v$ is not a constant but
        depends on spatial coordinates. We assume that $v$ approaches a constant value $v_0$ at spatial infinity,
\begin{equation}
v^2(\boldsymbol{x}) \longrightarrow v_0^2, \qquad r \equiv |\boldsymbol{x}| \longrightarrow \infty,
\end{equation}
where $\boldsymbol{x}$ refers the spatial coordinates, i.e., $\boldsymbol{x}=(x_1, x_2)$.
Then we can write
        \begin{align}
                v^{2}(\boldsymbol{x}) = v_0^2 + \sigma(\boldsymbol{x}), \label{030}
        \end{align}
where $\sigma(\boldsymbol{x})$ vanishes at spatial infinity.
One can consider such inhomogeneity to be due to impurities in the system or to be originated from larger theories.
Obviously, this explicitly breaks translation symmetry. As seen below, however, the BPS nature can be restored if another inhomogeneous term is added to the Lagrangian,
        \begin{align}
                \Delta\mathscr{L} = s \sigma(\boldsymbol{x}) B, \label{029}
        \end{align}
where $s$ is either $+1$ or $-1$ and $B$ is the magnetic field $B=\epsilon_{ij}\partial_i A_j$. This magnetic impurity term has been considered in the context
of abelian Higgs model as a supersymmetry-preserving impurity \cite{Hook:2013yda, Tong:2013iqa}. 
The Lagrangian density considered in the paper is then%
\footnote{Following \cite{Tong:2013iqa}, it is also possible to show that it 
originates from larger theories with two U(1) Chern-Simons 
fields~\cite{Kim:1993mh} where the impurities are realized as vorticies 
in the infinitely heavy limit.}
        \begin{align}
                \mathscr{L} = \frac{\kappa}2 \epsilon^{\mu\nu\rho} A_{\mu} \partial_{\nu} A_{\rho} - \overline{D_{\mu} \phi} D^{\mu} \phi
             - \frac{1}{\kappa^{2}} |\phi|^{2} (|\phi|^{2} - v^2(\boldsymbol{x}) )^{2} + s\sigma(\boldsymbol{x}) B
                , \label{010}
        \end{align}
where the last term comes from $\Delta\mathscr{L}$ in \eqref{029}.
This theory has been first introduced in \cite{Han:2015tga} 
and further studied in \cite{Bazeia:2024fgo}.

The energy of the theory reads
        \begin{equation}
                E = \int d^2x \Big[ |D_{0} \phi|^{2} + |D_{i} \phi|^{2} + \frac{1}{\kappa^{2}} |\phi|^{2} ( |\phi|^{2} - v^2(\boldsymbol{x}) )^{2} - s\sigma(\boldsymbol{x}) B \Big].
        \end{equation}
Now let us see how $\Delta\mathscr{L}$ helps in applying the Bogomolny trick to the energy.
It can be reshuffled to 
        \begin{equation}
E = \int d^{2} x\, \bigg\{ 
\left|D_0 \phi + i \frac{s}\kappa \phi(|\phi|^2-v^2(\boldsymbol{x})) \right|^2
+ |(D_{1} + i s D_{2}) \phi|^{2} 
+ s v^2(\boldsymbol{x}) B - s\sigma(\boldsymbol{x}) B\bigg\}
                ,\label{017}
        \end{equation}
up to a vanishing surface term. In obtaining this expression, we have used 
the relation $[D_{1} , D_{2}] = -i B$ and Gauss' law,
        \begin{equation}
                \kappa B = i (\bar{\phi} D^0 \phi - \overline{D^0 \phi} \phi) \equiv j^{0},  \label{gauss}
        \end{equation}
where we have introduced the U(1) current
        \begin{align}
                j^{\mu} = i (\bar{\phi} D^{\mu} \phi - \overline{D^{\mu} \phi} \phi)
                .\label{051}
        \end{align}
It is related to the electric field $E_i = F_{i0}$ through the equation of 
motion,
\begin{equation}
E_i = -\frac1\kappa \epsilon_{ik}j^k - s\partial_i \sigma.
\label{electric}
\end{equation}
Integrating the Gauss' law \eqref{gauss}, we can express the U(1) charge 
$Q_{\textrm{U(1)}}$ in terms of the magnetic flux $\Phi_B$,
\begin{equation} \label{qphi}
Q_{\textrm{U(1)}} = \kappa \Phi_B,
\end{equation}
which is a characteristic feature of the Chern-Simons gauge theory. 

Note that the inhomogeneous part in the last two terms in \eqref{017} are
cancelled. Then the energy is bounded from below by the magnetic flux or
the U(1) charge,
\begin{equation}\label{021}
E \ge s v_0^2 \int d^{2} x \, B = s v_0^2 \Phi_{B}
   = s \frac{v_0^2}\kappa Q_{\textrm{U(1)}}. 
\end{equation}
The bound is saturated if the following Bogomolny equations hold,
\begingroup
\allowdisplaybreaks
\begin{align}
(D_{1} + i s D_{2} ) \phi &= 0, \label{015}\\
D_0 \phi + i \frac{s}\kappa \phi(|\phi|^2-v^2(\boldsymbol{x})) &= 0. \label{0166}
\end{align}
\endgroup
With the help of the Gauss' law \eqref{gauss}, \eqref{0166} can also be written as
\begin{equation}
B + \frac{2s}{\kappa^2} |\phi|^2 (|\phi|^2 - v^2(\boldsymbol{x})) = 0. \label{016}
\end{equation}
It is straightforward to check that every static solution of these equations automatically satisfies the second-order Euler-Lagrange equations.

Once the sign $s$ of the inhomogeneous term \eqref{029} is fixed, so are the energy bound \eqref{021} as well as the Bogomolny equations.
In the usual homogeneous case where $\sigma(\boldsymbol{x})=0$, both $s=1$ and $s=-1$ are possible when completing the squares in the single theory, 
leading two separate energy bounds accordingly and hence $E \ge v_0^2 |\Phi_B|$. In the present case, however, we have only one energy bound because
it is tied to to sign in front of the inhomogeneous term. We will fix $s=1$ from now on without loss of generality. (One can obtain $s=-1$ case
under the parity transformation: $x_2 \rightarrow -x_2$ and $A_2 \rightarrow -A_2$.) 

It would be illuminating to consider energy-momentum tensor since the translation symmetry is broken.
For static configurations, the stress components $T_{ij}$ of symmetrized energy-momentum tensor can be written as
\begingroup
\allowdisplaybreaks
\begin{align}
T_{ij}  =&\, \frac{\kappa^2}{4|\phi|^2} 
    \Big[ B - \frac2{\kappa^2} |\phi|^2 (|\phi|^2 - v^2(\boldsymbol{x})) \Big] 
    \Big[ B + \frac2{\kappa^2} |\phi|^2 (|\phi|^2 - v^2(\boldsymbol{x})) \Big] 
    \delta_{ij} \nonumber\\
&\, + \frac14 \Big[ 
    \overline{(D_i + i\epsilon_{ik} D_k) \phi} (D_j - i \epsilon_{jl}D_l) \phi
   +(D_i + i\epsilon_{ik} D_k) \phi \overline{(D_j - i \epsilon_{jl}D_l) \phi}
    \nonumber\\
& \qquad + 
    \overline{(D_i - i\epsilon_{ik} D_k) \phi} (D_j + i \epsilon_{jl}D_l) \phi
   +(D_i - i\epsilon_{ik} D_k) \phi \overline{(D_j + i \epsilon_{jl}D_l) \phi}
    \Big]
        \label{028}
        ,
\end{align}
\endgroup
which vanishes on using the Bogomolny equations.
Therefore, the pressure density of the solutions vanish even in the presence
of the magnetic impurity. The conservation equation for the energy-momentum
tensor is modified to
\begin{equation}
\partial_{\mu} T^{\mu \nu} = 
\left[ B + \frac2{\kappa^2} |\phi|^2 (|\phi|^2 - v^2(\boldsymbol{x})) \right]
\partial^\nu \sigma(\boldsymbol{x}).
\label{tmui}
\end{equation}
Thus the momentum is not conserved, which is expected since the translation
symmetry is broken. Note that the right hand side is nothing but the Bogomolny
equation \eqref{016}. It is then zero for solutions of the Bogomolny 
equations, which is consistent with \eqref{028} since the solutions are
static.

A characteristic feature of Chern-Simons gauge theories is that BPS configurations can carry non-zero spin which is defined by
\begingroup
\allowdisplaybreaks
        \begin{align}
                J &= \int d^{2} x\, \epsilon^{ij} x_{i} T^0_{\ j} \nonumber \\
                  &= -\int d^{2} x\, \epsilon^{ij} x_{i} (\overline{D_{0} \phi} D_{j} \phi + \overline{D_{j} \phi} D_{0} \phi).
                \label{082}
        \end{align}
        \endgroup
Let us decompose the scalar field into 
the amplitude $|\phi|$ and the phase $\Omega$,
\begin{equation} \label{phiomega}
\phi = e^{i\Omega} |\phi|.
\end{equation}
For static configurations, we can rewrite $J$ as
\begingroup
\allowdisplaybreaks
\begin{align}
J &= -i \int d^{2} x\, \epsilon^{ij} x_{i} A_0 (\bar\phi D_j \phi - \phi \overline{D_j \phi}) \nonumber \\
  &= \kappa \int d^{2} x\, \epsilon^{ij} x_{i} \bar A_j B,
\label{angJ1}
\end{align}
\endgroup
where we used the Gauss' law \eqref{gauss} in the second line and
\begin{equation}
\bar A_i = A_i - \partial_i \Omega .
\label{barai}
\end{equation}
Now we impose the Bogomolny equation \eqref{015} (with $s=1$),
which can be expressed as
\begin{equation} \label{barAi}
\bar A_i = \epsilon_{ij} \partial^j \ln |\phi|.
\end{equation}
Then the angular momentum becomes
\begin{align}\label{angJ3}
        J = \frac{1}{\kappa} \int d^{2} x\, (|\phi|^{2} - v^{2} (\boldsymbol{x})) x_{i} \partial_{i} |\phi|^{2} ,
\end{align}
where we used \eqref{016}. These expressions will be used in later sections.

Though the Poincare symmetry is explicitly broken, BPS nature of the theory suggests that it still has a supersymmetric extension.
In fact, it is precisely given by the $\mathcal{N}=2$ supersymmetric homogeneous abelian CSH model \cite{Lee:1990it}
modified by $\Delta\mathscr{L}$ in \eqref{029} without further correction,
\begingroup
\allowdisplaybreaks
\begin{align}
\mathscr{L}_{\rm m} 
 =&\, \frac\kappa2 \epsilon^{\mu\nu\rho} A_{\mu} \partial_{\nu} A_{\rho}
 - \overline{D_{\mu} \phi} D^{\mu} \phi + i\bar{\psi} \gamma^{\mu} D_\mu \psi
 - \frac1{\kappa^2} |\phi|^{2} (|\phi|^{2} - v^2(\boldsymbol{x}))^2 \nonumber\\
 &\,-\frac{i}{\kappa} (3|\phi|^{2} - v^2(\boldsymbol{x})) \bar{\psi} \psi
    + \sigma(\boldsymbol{x}) B.
\label{008}
\end{align}
\endgroup
It is invariant up to a total derivative under the following supersymmetric transformation,
\begingroup
\allowdisplaybreaks
\begin{align}
&\delta \phi = i\bar{\eta} \psi, \nonumber\\
&\delta \bar{\phi} = i\bar{\psi} \eta, \nonumber\\
&\delta \psi = -\gamma^{\mu} \eta D_{\mu} \phi - \frac{1}{\kappa} (|\phi|^{2} - v^2(\boldsymbol{x})) \phi \eta ,\nonumber\\
&\delta \bar{\psi} = \overline{D_{\mu} \phi} \bar{\eta} \gamma^{\mu} - \frac{1}{\kappa} (|\phi|^{2} - v^2(\boldsymbol{x})) \bar{\phi} \bar{\eta} ,\nonumber\\
&\delta A_{\mu} = \frac{1}{\kappa} (\bar{\eta} \gamma_{\mu} \psi \bar{\phi} + \phi 
\bar{\psi} \gamma_{\mu} \eta), \label{005}
\end{align}
\endgroup
provided that the complex parameter $\eta$ satisfies the condition
        \begin{equation}
                \gamma^{1} \eta = -i \gamma^{2} \eta, \label{020}
        \end{equation}
where the gamma matrices are given by 
$\gamma^\mu = (i\sigma^2, \sigma^1, \sigma^3)$, $\mu=0,1,2$.
Thus the number of supersymmetry is reduced from $\mathcal{N}=2$ to $\mathcal{N}=1$ in inhomogeneous case.

With the condition \eqref{020}, the supersymmetric variation $\delta \psi$ 
of the fermion field in \eqref{005} can be written as,
\begin{equation}
\delta \psi = -\gamma^1 \eta (D_1 + i D_2)\phi 
        + i\left[ D_0 \phi + i\frac{1}{\kappa} (|\phi|^{2} - v^2(\boldsymbol{x})) \phi \right] \eta .
\end{equation}
It vanishes if
\begingroup
\allowdisplaybreaks 
\begin{align}
(D_1 + i D_2)\phi &=0, \notag \\
D_0 \phi + i\frac{1}{\kappa} \phi (|\phi|^{2} - v^2(\boldsymbol{x})) &= 0,
\end{align}
\endgroup
which are identical to \eqref{015} and \eqref{0166}, as it should be. 
The unbroken supercharges of the theory can be obtained by a standard procedure,
\begingroup
\allowdisplaybreaks
        \begin{align}
                Q &= \int d^2x\, \bigg\{ - i \Big[ D_{0} \phi + \frac{i}{\kappa} \phi (|\phi|^{2} - v^2(\boldsymbol{x}) ) + (D_{1} + i D_{2}) \phi \Big] \psi_{1}^{\dagger} 
                \nonumber\\
                & ~~~~~~~~~~~~~+ \big[ D_{0} \phi + \frac{i}{\kappa} \phi (|\phi|^{2} - v^2(\boldsymbol{x}) ) - (D_{1} + i D_{2} ) \phi\big] \psi_{2}^{\dagger} \bigg\} ,
                \nonumber\\
                Q^{\dagger} &= \int d^2x\, \bigg\{ i \Big[ \overline{D_{0} \phi} + \frac{i}{\kappa} \bar\phi ( |\phi|^2 - v^2(\boldsymbol{x}) ) + \overline{(D_{1} + i D_{2})\phi} \Big] \psi_{1} 
                \nonumber\\
                & ~~~~~~~~~~~~~ + \Big[ \overline{D_{0} \phi} + \frac{i}{\kappa} \bar{\phi} (|\phi|^{2} - v^2(\boldsymbol{x}) ) - \overline{(D_{1} + i D_{2} )\phi}\Big] \psi_{2} \bigg\}.
        \end{align}
\endgroup
They satisfy the superalgebra
\begingroup
\allowdisplaybreaks
        \begin{align}
        \{Q, Q \} &=\{Q^{\dagger}, Q^{\dagger} \}=0 ,\nonumber\\
        \{Q, Q^{\dagger} \} &= 2 \int d^2 x\, \Big[ | D_{0} \phi + \frac{i}{\kappa} \phi (|\phi|^{2} - v^2(\boldsymbol{x}) ) |^{2} + | (D_{1} + i D_{2})\phi |^{2} \Big] \nonumber\\
        &= 2 E - 2v_0^2 \Phi_B ,\label{125}
\end{align}
\endgroup
reproducing the energy bound \eqref{017}.

Although we will mainly consider the system with magnetic impurities in this
paper, it is also possible to add electric impurities or both to the system 
while keeping the BPS nature. We briefly discuss the case by considering
the off-shell formulation of its supersymmetric inhomogeneous model
in the appendix A.

Now let us come back to the Bogomolny equations.
It is well-known \cite{Jaffe:1980mj} that if $\phi$ is nonvanishing and satisfies \eqref{015} with $s=1$, then its zeros, if it has any, are isolated and
only positive vortex number is possible, i.e., $\Phi_B \ge 0$. Therefore, the theory can have BPS solutions only with positive vorticities, in contrast to
the homogeneous case where both BPS vortices and BPS antivortices exist.
Let $\boldsymbol{x}_a$ with $a=1,2,\ldots,n$ be the zeros of $\phi$. Eliminating the gauge field, it is straightforward to combine the two equations 
\eqref{015} and \eqref{016} into a single equation,
\begin{equation}
        \nabla^{2} \ln |\phi|^{2}= \frac{4}{\kappa^{2}}|\phi|^{2} [ |\phi|^{2} - v^2(\boldsymbol{x}) ]
                       + 4\pi \sum_{a=1}^n \delta(\boldsymbol{x}-\boldsymbol{x}_a).
                \label{026}
\end{equation}
In homogeneous case where $\sigma(\boldsymbol{x})=0$, i.e., $v^2(\boldsymbol{x}) = v_0^2$, solving \eqref{026} is equivalent to solving two Bogomolny 
equations. Then, the equation \eqref{026} is known to have topological as well as nontopological soliton solutions with both positive and negative vorticities \cite{Jackiw:1990pr}. 
For inhomogeneous case with $\sigma(\boldsymbol{x}) \neq 0$, however, we have only solutions with positive vorticities as mentioned above. Thus, only half of the solutions to \eqref{026} are true solutions in this case. 

\section{Inhomogeneous BPS Vacuum}

In usual field theory, the vacuum is given by constant field configurations since any physical variation in spacetime costs energy. If the translation symmetry is broken, however, there is no a priori reason for that. 
In this section, we will study the vacuum configurations of the inhomogeneous theory \eqref{010}. 
Recall that in the homogeneous theory with $\sigma = 0$, there are two vacua $\phi = 0$ and $|\phi|=v$ as discussed in the previous section. 
It is clear that the symmetric vacuum $\phi=0$ is still a vacuum solution 
even when $\sigma(\boldsymbol{x}) \neq 0$%
\footnote{In the symmetric vacuum with $\phi=0$, while the magnetic field 
vanishes, the electric field does not. This is because $\sigma(\boldsymbol{x})$
becomes a source for $E_i$ as seen in \eqref{electric}. Thus, $A_0$ is simply
given by $A_0 = -\sigma$ with $A_i=0$, which has no contribution to the 
energy.}. 
Now that $v(\boldsymbol{x})$ is position-dependent, however, the field configuration of the broken vacuum cannot be constant.
Moreover, naive $\phi = v(\boldsymbol{x})$ configuration would not minimize the energy.

In fact, since the energy is bounded from below by the magnetic flux as in \eqref{021}, any vacuum configuration should also satisfy Bogomolny equations \eqref{015} and \eqref{016} with vanishing magnetic flux $\Phi_B = 0$. Thus to obtain the vacuum configurations, we need to solve them in the $\Phi_B = 0$ sector.
It amounts to solve \eqref{026} without $\delta$-function terms in the right hand side, i.e.,
\begin{equation}
\nabla^{2} \ln |\phi|^{2}= \frac{4}{\kappa^{2}}|\phi|^{2} [ |\phi|^{2} - v^2(\boldsymbol{x}) ].
                \label{vacuumeq}
\end{equation}
It is evident that $\phi = v(\boldsymbol{x})$ cannot be a solution for any nontrivial $v(\boldsymbol{x})$ on $\mathbb{R}^2$ with the boundary condition 
$v(\boldsymbol{x}) \rightarrow v_0$ at spatial infinity%
\footnote{If we do not impose the condition that the inhomogeneity should vanish at spatial infinity, then $\phi=v(\boldsymbol{x})$ can be a solution if $v = |e^{f(z)}|$ for any nonsingular holomorphic function $f(z)$.}.
Nevertheless, it should be physically clear that there exists a nontrivial 
vacuum solution satisfying \eqref{vacuumeq} at least for a reasonable 
$v(\boldsymbol{x})$. It can actually be proved that this is indeed the
case by the same argument used in \cite{Han:2015tga}, details of which is 
beyond the scope of this paper.

In usual homogeneous theories, the vacuum is necessarily spinless. However, it is unclear whether this is true even for inhomogeneous cases, considering that BPS solutions of Chern-Simons gauge theories can have non-zero spins. Here we show that the vacuum is still spinless for arbitrary $v(\boldsymbol{x})$.
First, it is obvious that the unbroken vacuum is spinless since it is still
given by the vanishing configuration $\phi = 0$. To prove that the broken 
vacuum with the asymptotic behavior $\phi \rightarrow v_0$ remains spinless 
for any arbitrary inhomogeneous deformation, observe that the phase $\Omega$ 
in \eqref{phiomega} should be well-defined and regular everywhere for
the broken vacuum. Thus we have
$B = \epsilon^{ij} \partial_i A_j = \epsilon^{ij} \partial_i \bar A_j $.
Then, we can write \eqref{angJ1} as~\cite{Jackiw:1990pr, Kim:1992yz}
\begin{align}
J &= \kappa \int d^{2} x\, \partial_i \Big( \frac{1}{2} x^i \bar A_j \bar A^j - x_j \bar A^i \bar A^j \Big) + \kappa \int d^{2} x\, x_{i} \bar A^i \partial_j \bar A^j.
\label{angJ2}
\end{align}
But the last term vanishes on using \eqref{barAi}.
Moreover, the first term is a total derivative of a regular function for 
the broken vacuum solution. Then the first term becomes a boundary integral
over spatial infinity. Since $|\phi|$ approaches $v_0$ exponentially fast 
for the broken vacuum solution, \eqref{barAi} ensures that the boundary 
integral should vanish. This completes the proof.

As an explicit example, let us choose a rotationally symmetric 
$v^2(\boldsymbol{x}) = v_0^2 + \sigma(\boldsymbol{x})$ with 
\begin{equation}
\sigma(\boldsymbol{x}) = - \beta v_0^2 e^{-\alpha^2 m^2 r^2},
\label{sigmax}
\end{equation}
where
\begin{equation}
m = \frac{2v_0^2}\kappa
\end{equation}
is the scalar mass \eqref{047} in $r\rightarrow\infty$ limit. Then,
there is a Gaussian dip (or bump) of size $1/\alpha m$ at the origin and
its depth is controlled by the parameter $\beta$. Note in particular that,
for $\beta = 1$, the naive ``vacuum expectation value'' $v(\boldsymbol{x})$ 
of the broken vacuum vanishes at the origin. To find the broken vacuum 
solution  of \eqref{026} deformed from $\phi = v_0$, we adopt a rotationally 
symmetric ansatz $\phi = |\phi(r)|$ without any phase factor. Being a broken
vacuum, it should have no zero and the appropriate boundary conditions are
\begingroup
\allowdisplaybreaks
\begin{align}
\lim_{r \rightarrow \infty} \phi &= v_0, \nonumber \\
\lim_{r \rightarrow 0} \frac{d\phi}{dr} &= 0.
\end{align}
\endgroup
Then \eqref{026} becomes
\begingroup
\allowdisplaybreaks
\begin{align}
\frac{d^{2} \ln |\phi|^2}{dr^{2}} + \frac{1}{r} \frac{d\ln |\phi|^2}{dr}
 = \frac4{\kappa^2}|\phi|^2[|\phi|^2-v_0^2(1 - \beta e^{-\alpha^2 m^2 r^2}) ]
        .\label{049}
\end{align}
\endgroup

\begin{figure}[H]
                \centering
                \subfigure[ ]{
                        \includegraphics[page=1,width=0.45\textwidth]{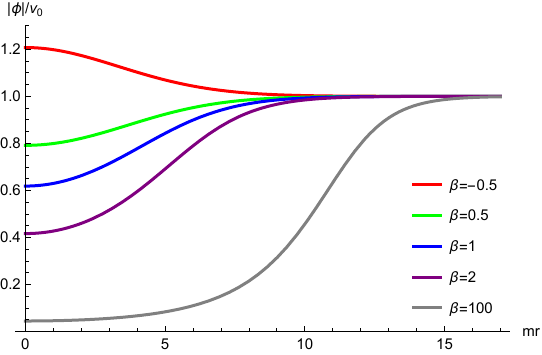}
                }
                \subfigure[ ]{
                        \includegraphics[page=2,width=0.45\textwidth]{plotvac.pdf}
                }
\caption{(a) Amplitudes of the scalar field of the inhomogeneous broken vacuum
for a Gaussian impurity \eqref{sigmax} with $\alpha^{-2}=20$ and various 
$\beta$. 
(b) Energy densities $T_{00}$ of the vacuum in unit of $\kappa m^3$.}
\label{fig:solT00vacbeta}
        \end{figure}

Solving \eqref{049} near the origin, we get
\begin{align} \label{vorigin}
|\phi| &\approx a v_0 \left\{ 1 + \frac{a^2}8 (a^2 -1 + \beta) (mr)^2 \right.
        \nonumber \\
&\qquad \left.
                + \frac{a^{2}}{128} \big[ a^{2} (a^{2} - 1) (3 a^{2} - 2) - (4 a^{2} - 5 a^{4} + 4 \alpha^{2}) \beta +  2 a^{2} \beta^{2} \big] (mr)^{4} + \cdots \right\},
\end{align}
where $a$ is a constant. At large distances, the asymptotic behavior of the 
solution is independent of the inhomogeneous term,
\begin{equation} \label{asymp}
|\phi| \approx v_0 [ 1 - b K_{0} (m r) ] ,  
\end{equation}
where $b$ is some constant. Then there should be some definite value
$a = a_v$ which allows $|\phi|$ to smoothly connect to \eqref{asymp} with some 
$b$ asymptotically. It is worth noting that if $\beta = 0$ in \eqref{vorigin}, 
we recover the constant broken vacuum solution $ |\phi| = v_0 $ with $a=1$ 
in the homogeneous theory.
Figure \ref{fig:solT00vacbeta} shows typical profiles of the inhomogeneous
BPS vacuum configuration $\phi$ and the corresponding energy density $T_{00}$ 
obtained by numerical works for a fixed $\alpha^{-2} = 20$ 
and various $\beta$. The size of the region where $\phi$ is substantially
different from the asymptotic value $v_0$ is determined by $\alpha$ and
is not much dependent on the mass scale \eqref{047}, since it is due to the
inhomogeneous term \eqref{029} added to the theory.
As $\beta$ increases, the Gaussian dip at the origin gets deeper and the
profile of $\phi$ near the origin moves towards the symmetric vacuum $\phi = 0$
away from the asymptotic value ($|\phi| = v_0$). 
For negative $\beta$, the solution starts with a value
larger than $v_0$ so that it decreases to $v_0$ asymptotically.

Note that $v^2(0)$ corresponds
to zero for $\beta = 1 $ and $v^2(0)$ becomes negative for $\beta >1$. 
Thus, for $\beta >1$, $|\phi^2| = v^2(\boldsymbol{x})$ can no 
longer be a minimum of the potential around the origin. Nevertheless, we see 
that there is a broken vacuum solution which is completely smooth and 
well-defined. It is worth mentioning that the energy of the solution 
is indeed zero in an interesting way. Although the energy density $T_{00}$
near the origin is negative, it becomes positive in the ring-shaped region
encircling the negative-energy part so that the total integral of the energy 
density makes exactly zero, as it should be.

\begin{figure}[H]
                \vspace{-3mm}
                \begin{center}
                        \includegraphics[page=4,width=0.55\textwidth]{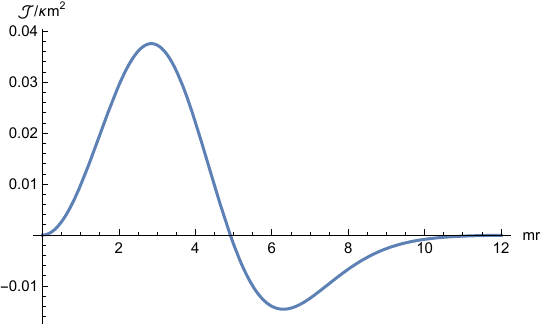}
                \end{center}
\caption{Angular momentum density $\mathcal{J}$ of the inhomogeneous broken
vacuum in unit of $\kappa m^2$ for a Gaussian impurity \eqref{sigmax} with 
$\alpha^{-2}=20$ and $\beta=1$.}
\label{fig:angJ}
        \end{figure}

One can obtain a similar behavior for the angular momentum $J$ in \eqref{angJ1}.
Namely, the angular momentum density $\mathcal{J}=\epsilon^{ij}x_i T^{0j}$
is locally nonzero but its integral over the space is zero, in accordance
with what we have shown above. This can be seen from the expression 
\eqref{angJ3}. Since $|\phi|$ is an increasing function of $r$ for the vacuum
solution, the sign of $\mathcal{J}$ is completely determined by the factor
$|\phi|^2-v^2(\boldsymbol{x})$. Thus $\mathcal{J}$ 
should be positive near the origin but becomes negative for large $r$. 
In Figure \ref{fig:angJ}, we plot $\mathcal{J}$ for $\alpha^{-2} = 20$ 
and $\beta = 1$. At the center $\mathcal{J}$ vanishes. Then, it is surrounded 
by a ring with positive $\mathcal{J}$, which is, in turn, compensated by 
negative $\mathcal{J}$ in the outer region, as it should be.
Therefore, the vacuum consists of regions rotating in
different directions. Comparing with the energy density $T_{00}$ in 
Figure \ref{fig:solT00vacbeta}, we see that regions with positive/negative 
$T_{00}$ do not coincide with those with positive/negative $\mathcal{J}$.
Considering that the inhomogeneous source $\sigma(\boldsymbol{x})$ does
not directly contribute to $T_{0i}$, it is rather intriguing how the fields 
conspire to keep the vacuum spinless by locally adjusting the angular momentum
density, while maintaining zero energy at the same time. It should be due to 
a nontrivial role of the Chern-Simons gauge field. In fact this kind of 
phenomenon does not occur in the inhomogeneous abelian Higgs model \cite{Kim:2024gfn}.

\begin{figure}[H]
                \centering
                \subfigure[]{
                        \includegraphics[page=5,width=0.45\textwidth]{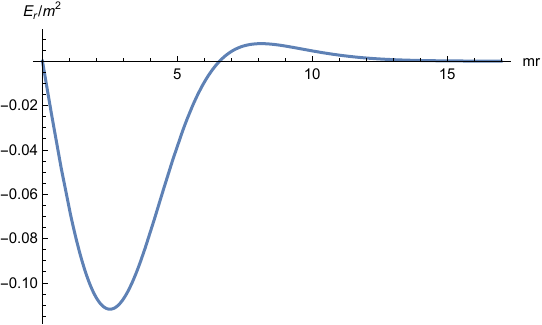}
                }
                \subfigure[]{
                        \includegraphics[page=3,width=0.45\textwidth]{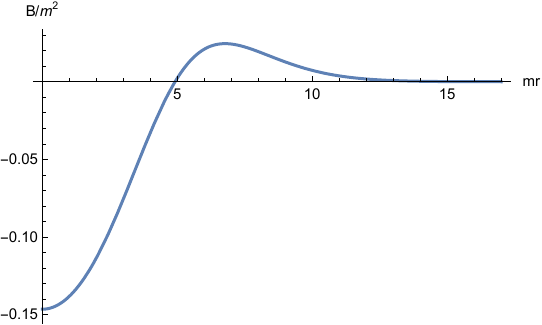}
                }
\caption{(a) The electric field and (b) the magnetic field of the 
inhomogeneous broken vacuum in unit of $m^2$ for a Gaussian 
impurity \eqref{sigmax} with $\alpha^{-2}=20$ and $\beta=1$.}
\label{fig:EBvac}
        \end{figure}

The electric and the magnetic fields are also nonvanishing, which can be
calculated from \eqref{electric} and \eqref{gauss}. The electric field
has nonvanishing radial component $E_r$,
\begin{equation}
E_r = \frac1\kappa \frac{d}{dr}(|\phi|^2 - v^2) .
\end{equation}
Similarly, the magnetic field is nonzero since the vacuum solution is not 
simply given by $|\phi|^2 = v(\boldsymbol{x})^2$, although the entire magnetic 
flux $\Phi_B$ as well as the charge $Q_{\textrm{U(1)}}$ should be zero for the 
vacuum solution. We plot the these fields in Figure \ref{fig:EBvac}. 

Now we consider the $\delta$-function limit of
the impurity~\eqref{sigmax}. With $\beta = 4\alpha\eta$, the inhomogeneous
function $\sigma$ in \eqref{sigmax} is rewritten as
\begin{align} \label{delta}
\sigma(r) &= -4 \eta \alpha v_0^2 e^{-\alpha m^2 r^2} \nonumber \\
 &\equiv -\frac{\pi\eta\kappa^2}{v_0^2} \delta_{\alpha}(r),
\end{align}
where $\delta_{\alpha}(r)$ becomes the two-dimensional delta function
$\delta(\boldsymbol{x})$ in $\alpha \rightarrow \infty$ limit for fixed
$\eta$. The vacuum equation \eqref{vacuumeq} then leads to
\begin{equation}
\nabla^{2} \ln |\phi|^{2}= \frac{4}{\kappa^{2}}|\phi|^{2} ( |\phi|^2 - v_0^2)
+\frac{4\pi\eta}{v_0^2}|\phi(\boldsymbol{x})|^2 \delta_\alpha(r) .
                \label{deltavaceq}
\end{equation}
Note that $|\phi(\boldsymbol{x})|^2$ is multiplied in front of the impurity
function $\delta_\alpha(r)$, which is a characteristic feature of the
BPS CSH model having the sextic potential \eqref{011}.  Without this factor,
the delta-function limit of the last term in \eqref{deltavaceq} is similar
to the delta-function term in \eqref{026}, which comes from the vortex points 
where the scalar field $\phi$ vanishes. Then, near the impurity point $r=0$,
$\phi$ would tend to zero with some power of $r$ which depends on the 
coefficient of the delta function%
\footnote{This is exactly the case in the inhomogeneous abelian Higgs model 
where the dynamics is governed by the Maxwell term~\cite{Kim:2024gfn}. 
In this model, the vacuum equation corresponding to \eqref{deltavaceq} is
\begin{equation}
\nabla^{2} \ln |\phi|^{2}
= 2g^{2} ( |\phi|^{2} - v_0^{2} ) + 4\pi\eta\delta_\alpha(\boldsymbol{x}) ,
\end{equation}
where $g$ is the coupling constant of the model. 
Thus, $\phi \sim r^\eta$ near the origin.}.%
(See section 4.) However, due to the factor $|\phi|^2$ in front of the delta 
function, this cannot happen because the vanishing $|\phi(0)|$ would kill
the delta function itself. So one should not naively assume that
the delta-function impurity would induce a zero of $\phi$.

\begin{figure}[H]
                \centering
                \subfigure[]{
                        \includegraphics[page=1,width=0.45\textwidth]{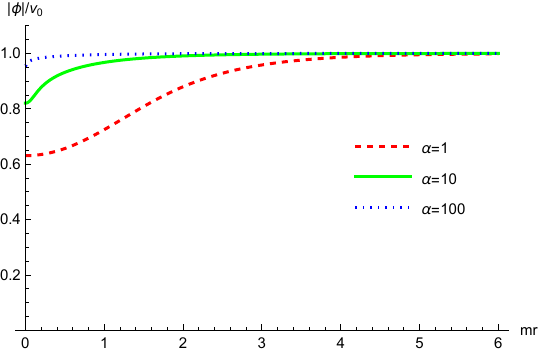} }
                \subfigure[]{
                        \includegraphics[page=2,width=0.45\textwidth]{deltaImCSH.pdf}
                }
\caption{(a) Amplitudes of the scalar field of the inhomogeneous broken vacuum
for different $\alpha$ and $\beta=4\alpha\eta$ with $\eta=1$ fixed.
(b) Energy densities $T_{00}$ of the vacuum in unit of $\kappa m^3$.}
\label{fig:delta}
        \end{figure}

To see the effect of the delta-function impurity, we obtain numerical
solutions of the vacuum by changing the size parameter $\alpha$ and the depth
parameter $\beta = 4\alpha\eta$ for fixed $\eta$.
Figure \ref{fig:delta} shows that, as we increase $\alpha$, the dip of
$|\phi|$ at the origin becomes narrower, as expected, but also more shallow
at the same time. It seems clear from the figure that,
as $\alpha \rightarrow \infty$, the vacuum solution simply becomes the 
constant configuration $|\phi|=v_0$ throughout the space including the origin.
Although the scalar field $\phi$ does not directly show the effect of the 
delta function impurity, it appears in the the energy density as shown 
in Figure~\ref{fig:delta} where delta function term
can be identified. This can also be seen from the Bogomolny equation
\eqref{016}. With $|\phi| \simeq v_0$, the magnetic field behaves as
\begin{equation}
B \simeq \frac{2 v_0^2}{\kappa^2}\sigma(\boldsymbol{x}) 
 = 2\pi \eta \delta_\alpha(r).
\end{equation}
 Nevertheless, the total magnetic flux as well as the total energy should 
remain zero since this is the vacuum solution. The localized flux and energy 
at the origin are compensated by the slight excess of the flux and energy 
in the rest of the space, which eventually go to zero in the delta-function 
limit.
 
There is a subtle point in this result. One may argue that the constant 
solution $|\phi| = v_0$  seem to contradict 
the existence of the delta function in \eqref{deltavaceq}, since it does not 
vanish with nonzero $|\phi(0)|$ and should have some effect on the solution. 
What we really find here is that the delta-function limit of the solution
is not well-defined at the origin; it becomes singular 
and the value of $\phi(0)$ and/or $\phi'(0)$ depends on the limiting
procedure. With the Gaussian regularization of the delta function adopted
here, $\phi'(0)$ becomes ill-defined, although $\phi(0)$ itself goes to $v_0$.
On the other hand, it can be shown that a different result is obtained
if a different regularization is used. We verify this explicitly in
Appendix B. The difference, however, exists just at the single point, which
should not change any physical quantities. 

\section{BPS Solitons with Inhomogeneous Mass}
We are now going to discuss solutions of Bogomolny equations \eqref{015} and
\eqref{016} with nonzero energies. Let us first recall the solutions
in the usual homogeneous case with $\sigma(\boldsymbol{x}) = 0$. Thanks to
the sextic potential with degenerate vacua $|\phi|=0$ and $|\phi|=v_0$,
the equations support rich soliton spectrum such as topological vortices, 
nontopological solitons (Q-balls), and nontopological vortices 
(Q-vortices)~\cite{Jackiw:1990pr, Hong:1990yh, Jackiw:1990aw}.
Since the inhomogeneity considered here is local and does not change the 
essential vacuum structure as seen in the previous section, we expect that
the same type of solitons exist in inhomogeneous theories. 
In \cite{Han:2015tga}, it was argued that topological vortices with positive 
vorticities exist if the source $\sigma(\boldsymbol{x})$ is square-integrable.

In this section we first study rotationally symmetric solutions
with Gaussian-type inhomogeneity \eqref{sigmax}. In particular, we mainly
consider the case $\beta=1$ so that $v^2(\boldsymbol{x}) = 0$ at the origin
and could affect the behavior of solutions near the origin. Then at the end of 
the section, we discuss solutions without rotational symmetry.

The relevant ansatz would be
\begingroup
\allowdisplaybreaks
\begin{align}
\phi &= |\phi(r)| e^{in\theta}, \nonumber \\
A^{i} &= -\epsilon^{ij} x_{j} \frac{A_{\theta} (r)}{r^{2}}.
\label{061}
\end{align}
\endgroup
Here $n$ is the vorticity of the solution and should be a nonnegative integer
as discussed at the end of Section 2. For regular solutions, we should have 
$A_\theta(0) = 0$ and $|\phi(r)| \sim r^n$ as $r \rightarrow 0$.
This ansatz is suitable for obtaining
solutions with $n$ zeros of $\phi$ all located at the origin.
Then the Bogomolny equations \eqref{015} and \eqref{016} become
\begingroup
\allowdisplaybreaks
\begin{align}
\frac{d|\phi|}{dr} &= \frac1r (n - A_{\theta}) |\phi| ,\label{018}\\
\frac{1}{r} \frac{dA_{\theta}}{d r} 
       &= \frac{2}{\kappa^2} |\phi|^2[v_0^2 + \sigma(r)-|\phi|^2] .\label{094}
\end{align}
\endgroup
Combining the two equations, we can also obtain a single second-order 
equation \eqref{049} for $r>0$. 

To find finite energy solutions, we need to impose the boundary conditions 
that $|\phi|$ approaches the vacuum value as $r \rightarrow \infty$,
\begingroup
\allowdisplaybreaks
\begin{align}
|\phi(r)| &\sim v_0,
&&A_\theta(r) \sim n \qquad &&\textrm{(broken phase),} \nonumber \\
|\phi(r)| &\sim  r^{-\varepsilon},
&&A_\theta(r) \sim n+\varepsilon \qquad &&\textrm{(symmetric phase),}
\end{align}
\endgroup
where $\varepsilon$ is some positive constant and we have used 
\eqref{018} to get boundary conditions for $A_\theta$. With these ansatz,
the magnetic flux $\Phi_B$ and the angular momentum $J$ in \eqref{angJ2} can
be determined as
\begingroup
\allowdisplaybreaks
\begin{align} \label{phij}
\Phi_B &= 2\pi [A_\theta(\infty) - A_\theta(0) ]
                   = 2 \pi (n + \varepsilon), \nonumber \\
J &= \pi \kappa [(n-A_\theta(\infty))^2 - (n-A_\theta(0))^2 ]
   = \pi \kappa (\varepsilon^2 - n^2),
\end{align}
\endgroup
where we put $\varepsilon = 0$ for broken phase. Note that these values are
the same as those of the homogeneous case \cite{Jackiw:1990pr},
independent of the details of the inhomogeneity $\sigma(r)$, since they are
completely determined by the boundary conditions.

It is convenient to classify the solutions by asymptotic values of 
$|\phi|$ and also by the vorticity $n$.
\begin{enumerate}
\item $|\phi(\infty)| = v_0$ and $n=0$: This is nothing but the vacuum solution
discussed in section 3.

\item $|\phi(\infty)| = v_0$ and $n \neq 0$: These are topological vortex
solutions~\cite{Hong:1990yh,Jackiw:1990aw,Han:2015tga} deformed in the 
inhomogeneous background $\sigma(\boldsymbol{x})$.

\item $|\phi(\infty)| = 0$ and $n=0$: Other than the trivial symmetric vacuum
$\phi=0$, these are nontopological solitons (Q-balls)~\cite{Jackiw:1990pr} 
deformed by $\sigma(\boldsymbol{x})$.

\item $|\phi(\infty)| = 0$ and $n \neq 0$: These solutions are nontopological
vortices and are called Q-vortices~\cite{Jackiw:1990pr} deformed by 
$\sigma(\boldsymbol{x})$.
\end{enumerate}

In the following we discuss each solution in detail.

\subsection{Topological vortices}
As in the homogeneous case $\sigma(\boldsymbol{x})=0$, 
$|\phi(\infty)| = v_0$ and $n \neq 0$ correspond to topological vortex
solutions~\cite{Han:2015tga}. At large distances, the asymptotic behaviors 
of the solutions are the same as \eqref{asymp}.
Near the origin, we obtain
\begingroup
\allowdisplaybreaks
\begin{align}
|\phi| &\approx a\, v_0\, (mr)^n \left[ 1
        - \frac{a^2 (\alpha^2 - a^2 \delta_{n1})}{8(n+2)^2} (mr)^{2n+4}
        + \frac{a^2 (\alpha^4 + 2 a^2 \delta_{n2})}{16(n+3)^2} (mr)^{2n+6}
        + \cdots \right], \nonumber \\
A_{\theta} &\approx 
        \frac{a^2 (\alpha^2 - a^2 \delta_{n1})}{4(n+2)} (mr)^{2n+4}
        - \frac{a^2 (\alpha^4 + 2 a^2 \delta_{n2})}{8(n+3)} (mr)^{2n+6}
        + \cdots
\label{054}
\end{align}
\endgroup
for some constant $a$ with $\beta=1$. Except the leading term $r^n$ in 
$|\phi|$, the expansion \eqref{054} is completely different from 
the homogeneous case \cite{Hong:1990yh,Jackiw:1990aw} which corresponds
to $\alpha \rightarrow \infty$ limit. This is because the source term 
\eqref{sigmax} with $\beta=1$ cancels $v_0$ at the origin so that $v(0)=0$. 
Nevertheless, topological vortex solutions can be obtained by adjusting the 
parameters $a$ and $b$. In fact, if the parameter $a$ is chosen too large, 
$|\phi|$ reaches $v_0$ at some finite $r$ and then diverges. If $a$ is chosen 
too small, $|\phi|$ goes to zero asymptotically. Then, there should be a 
unique value $a=a_n^v$ for each vorticity $n$ which satisfies the boundary 
condition $|\phi(\infty)| = v_0$. 

\begin{figure}[H]
\centering
\subfigure[]{
\includegraphics[page=1,width=0.45\textwidth]{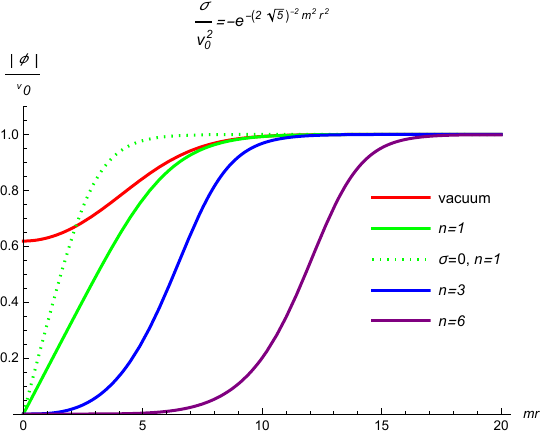}
5}
\subfigure[]{
\includegraphics[page=2,width=0.45\textwidth]{plottopovor.pdf}
}
\caption{(a) Amplitudes of the scalar field of topological vortices
for a Gaussian impurity \eqref{sigmax} with $\alpha^{-2}=20$ and $\beta=1$.
As a reference, $n=1$ vortex with $\sigma=0$ is also plotted as a dotted line.
(b) Energy densities $T_{00}$ of the topological vortices in unit of 
$\kappa m^3$.}
\label{fig:topovor1}
\end{figure}

In Figure \ref{fig:topovor1},
we plot numerical solutions for $n=1,3$ and 6 as illustrations. We also compare 
them with the vacuum solution discussed in section 3 as well as with $n=1$
solution of the homogeneous theory ($\sigma=0$). We can easily see the
effect of $\sigma$ in two $n=1$ solutions. The energy densities $T_{00}$ of 
the vortex solutions 
vanish at the origin except $n=1$ case. Apparently, this is the same behavior 
as in the homogeneous case. Note, however, that $T_{00} <0$ for the vacuum
solution. Thus, in this sense, the vortex contribution to energy relative 
to the vacuum is nonzero positive near the origin, which is to be contrasted 
with the homogeneous case where $T_{00}=0$ at the origin for both the vacuum
and the vortices with $n>1$.

\begin{figure}[H]
                        \centering
                        \subfigure[]{
                                \includegraphics[page=3,width=0.45\textwidth]{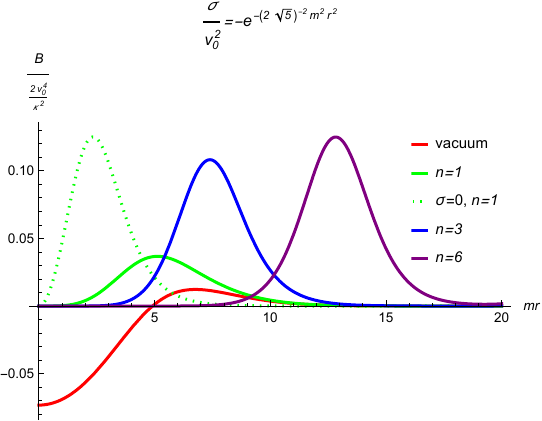}
                        }
                        \subfigure[]{
                                \includegraphics[page=4,width=0.45\textwidth]{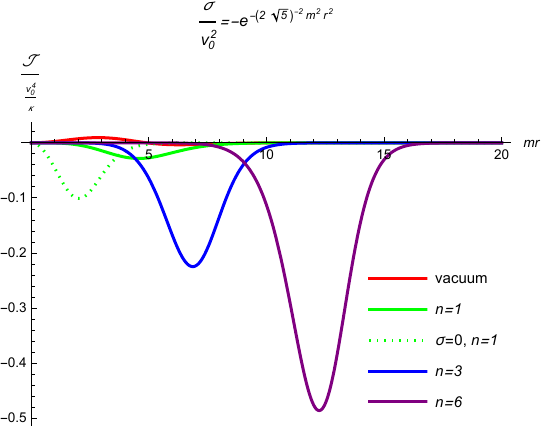}
                        }
\caption{(a) Magnetic fields of topological vortices in unit of 
$m^2$ for a Gaussian impurity \eqref{sigmax} with 
$\alpha^{-2}=20$ and $\beta=1$.
As a reference, $n=1$ vortex with $\sigma=0$ is also plotted as a dotted line.
(b) Angular Momentum densities of topological vortices in unit of 
$\kappa m^2$.}
\label{fig:solT00topovor}
                \end{figure}

We also plot the magnetic field and the angular momentum density in 
Figure \ref{fig:solT00topovor}.
Although the magnetic fields of the
vortices vanish at the origin and appear ring-shaped, the net vortex 
contributions do not, as the magnetic field of the vacuum is negative near
the origin, which is again different from the homogeneous system. Therefore,
one may say that the vortices restore the energy and the magnetic field at the 
impurity position which are depleted and expelled outwards by the impurity
$\sigma(\boldsymbol{x})$.
                
The objects are spinning with negative angular momentum $J=-\pi \kappa n^2$.
As in section 3, we can understand the sign from $\mathcal{J}$ in 
\eqref{angJ3}, where the factor $|\phi|^2 - v^2(\boldsymbol{x})$ is always 
negative for topological vortices while $d|\phi|/dr$ is positive.

We conclude this subsection with a brief comment on the delta function
impurity considered at the end of section 3. Since vortex solutions
behave as $|\phi| \sim r^n$ near the origin, the presence of the impurity term 
$|\phi|^2 \delta(\boldsymbol{x})$ should not alter the solutions of
the homogeneous theory as long as the nonzero vorticity $n$ is taken into
account. This is true for both topological and nontopological vortices discussed
in subsection 4.3. The only possible exception is the Q-ball solutions 
in the next subsection, but this case is exactly the same as the vacuum 
case.
 
\subsection{Nontopological solitons}
When $|\phi(\infty)|=0$ with $n=0$, an obvious solution is the trivial symmetric
vacuum solution $\phi=0$ as discussed in section 3. There exist, however, 
nontopological soliton solutions characterized by the value of the magnetic 
flux $\Phi_B = 2\pi \varepsilon$ as shown in \eqref{phij}. 
By the Gauss' law, this object also has a nonzero charge 
$Q_{\text{U}(1)} = \kappa \Phi_B = 2\pi \kappa \varepsilon$,
and is also called a Q-ball.

\begin{figure}[H]
                        \begin{center}
                                \includegraphics[page=1,width=0.55\textwidth]{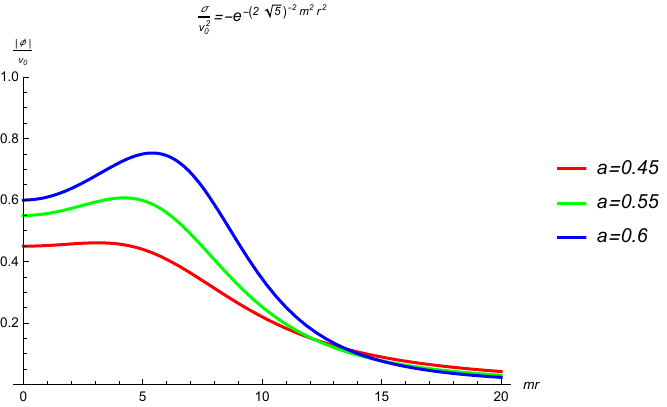}
                        \end{center}
\caption{Amplitudes of the scalar field of nontopological solitons 
(Q-balls) for various $a$ in \eqref{054} for a Gaussian impurity 
\eqref{sigmax} with $\alpha^{-2}=20$ and $\beta=1$.}
\label{fig:solnontoposol}
                \end{figure}

\begin{figure}[H]
                        \begin{center}
        \includegraphics[page=3, width=0.55\textwidth]{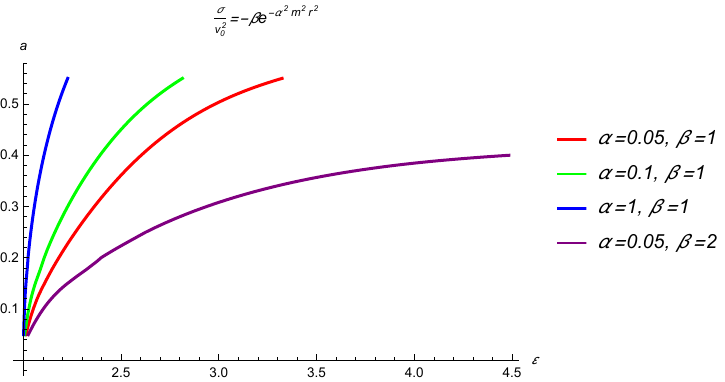}
                        \end{center}
\caption{Scalar amplitude at the origin $a$ as a function of 
the decay exponent $\varepsilon$ for a Gaussian impurity 
\eqref{sigmax} with various $\alpha$ and $\beta$. Note that the lower
limits of $\varepsilon$ are almost 2 which is the Liouville limit in the 
homogeneous theory.}
\label{fig:nontoposolepsilon}
                \end{figure}

Since $|\phi(0)|$ is nonvanishing, the behavior of the solution near the origin
is the same as the broken vacuum discussed in section 3, i.e., it is given
by \eqref{vorigin}. The constant $a$, however, should be 
smaller than $a_v$ in this case so that $|\phi|$ decays to zero asymptotically. 
Any nonzero positive $a$ smaller than $a_v$ should yield a Q-ball solution.
At large distances, the impurity $\sigma$ decays exponentially and should 
have no effect on the power-law behavior as $r \rightarrow \infty$. 
Thus it is identical to the homogeneous case in \cite{Jackiw:1990pr}, namely,
\begingroup
\allowdisplaybreaks
\begin{align} \label{rlarge}
|\phi| &\approx \frac{b}{(mr)^\varepsilon} \left[ 1
 - \frac{b^2}{8(\varepsilon-1)^2 (mr)^{2\varepsilon-2}} + \cdots \right], \nonumber \\
A_{\theta} &\approx \varepsilon -
        \frac{b^2}{4(\varepsilon-1) (mr)^{2\varepsilon-2}} + \cdots,
\end{align}
\endgroup
for some constants $b$ and $\varepsilon$ which will be determined by $a$.
Recall that the coefficient of the quadratic term in \eqref{vorigin} is
always positive with $\beta=1$ unlike the homogeneous case where $\beta=0$.
Then $|\phi|$ should increase near the origin as a function of $r$ and then 
decreases to zero as $r$ goes to infinity.
Such behaviors are clearly seen in Figure \ref{fig:solnontoposol} which is
obtained numerically. 
In this figure, we can also notice that solutions
with large $a$ corresponds to large $\varepsilon$ or large $\Phi_B$.
We have numerically obtained the relation between $\varepsilon$ and $a$
which is shown in Figure \ref{fig:nontoposolepsilon}.

In the homogeneous case, it is known \cite{Jackiw:1990pr} 
that $\varepsilon>2$ with $\varepsilon=2$ being
small-$|\phi|$ limit, where the Bogomolny equation \eqref{026} becomes the
Liouville equation since $|\phi|^2$ can be neglected in the expression
$|\phi|^2 -v^2$ for constant $v$. In the present case, this approximation 
is no longer valid since $v^2(\boldsymbol{x})$ can be smaller than $|\phi|^2$
near the origin for $\beta \ge 1$. Nevertheless, it is clear that there should
be a lower bound for $\varepsilon$ because $\varepsilon > 1$ for the energy to
be finite. We tried to obtain the bound numerically for various $\alpha$ and 
$\beta$, expecting that it would vary as a function of them. 
We find that, for $\beta>0$, the lower bound does not change much from
$\varepsilon=2$ as seen in Figure \ref{fig:nontoposolepsilon}, while it tends 
to be less than 2 for $\beta<0$. The reason behind this phenomenon is not 
clear to us at the moment.
We leave this issue as a future problem.

\begin{figure}[H]
        \centering
        \subfigure[]{
                \includegraphics[page=5, width=0.45\textwidth]{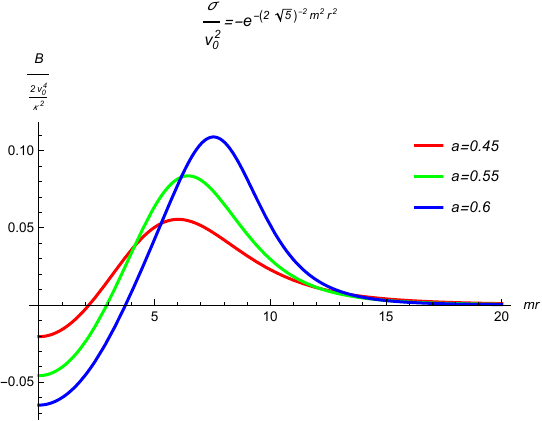}
        }
        \subfigure[]{
                \includegraphics[page=4, width=0.45\textwidth]{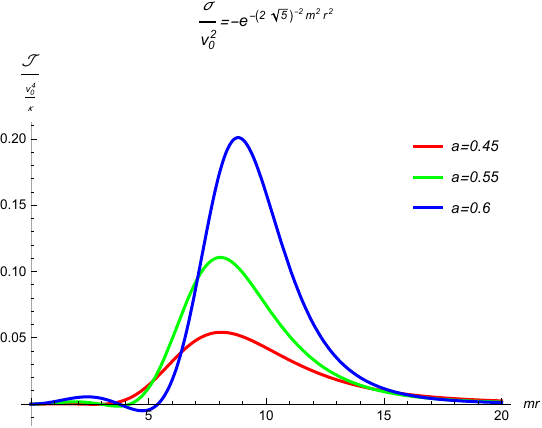}
        }
\caption{(a) Magnetic fields of nontopological solitons in unit of 
$m^2$ for a Gaussian impurity \eqref{sigmax} with 
$\alpha^{-2}=20$ and $\beta=1$.
(b) Angular Momentum densities of nontopological solitons in unit of 
$\kappa m^2$.}
\label{nontopbj}
\end{figure}

The magnetic flux and the angular momentum are given by 
$\Phi_B = 2\pi\varepsilon$ and $J = \pi\kappa\varepsilon^2$, respectively.
Then, they have also the corresponding lower bounds irrespective of $\sigma$.
In Figure \ref{nontopbj}, we plot the magnetic fields and the angular momentum 
densities for some typical values of $a$ with $\alpha^{-2} = 20$ and 
$\beta = 1$. Note that the angular momentum density $\mathcal{J}$ is mostly 
positive, which can be understood from the expression \eqref{angJ3} 
since $|\phi|$ decays to zero for most of the region.

Since the solutions are in the topologically trivial sector, one has to
worry about their stability against decaying into other excitations including perturbative charged particles. From
\eqref{021}, we see that the energy is given by
\begin{equation} \label{qenergy}
E = m_\phi |Q_{\text{U}(1)}|,
\end{equation}
where $m_\phi$ is the scalar mass \eqref{043} in the symmetric phase. Then the
energy per unit charge is the same as that of the elementary excitation.
It implies that the nontopological soliton is marginally stable, just as in the
homogeneous case \cite{Jackiw:1990pr}.

\subsection{Nontopological vortices}
We get nontopological vortices or Q-vortices with asymptotically decaying
$n\neq 0$ solutions. They may be considered as hybrids of previous two 
cases, i.e., nontopological solitons with vortices embedded at the center.
The series expansion near the origin is the same as \eqref{054} where 
the constant $a$ should now be smaller than $a_n^v$ for each $n$
which is the value for topological vortices. The asymptotic behavior is
again given by \eqref{rlarge}.

\begin{figure}[H]
\subfigure[]{
\includegraphics[page=1, width=0.45\textwidth]{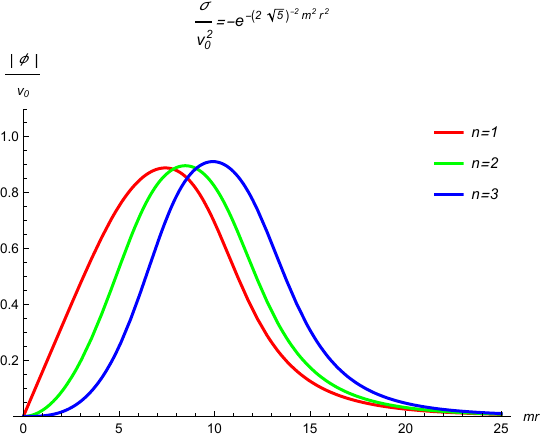}
}
\subfigure[]{
\includegraphics[page=2, width=0.45\textwidth]{plotnontopovor.pdf}
}
\caption{(a) Amplitudes of the scalar field of nontopological vortices
for a Gaussian impurity \eqref{sigmax} with $\alpha^{-2}=20$ and $\beta=1$.
(b) Energy densities $T_{00}$ of the nontopological vortices in unit of 
$\kappa m^3$.}
\label{fig:nontopvor}
\end{figure}

In Figure \ref{fig:nontopvor}, we plot numerical solutions for $n=1,2$ and 3.
Double-ring shape of the energy density $T_{00}$ shows that these solutions
are hybrids of vortices and Q-balls of magnetic flux $\Phi_{B} = 2\pi (n+\varepsilon)$.
The solutions are also marginally stable against decaying into elementary
excitations with energy \eqref{qenergy}.
                
\begin{figure}[H]
\begin{center}
        \includegraphics[page=4, width=0.55\textwidth]{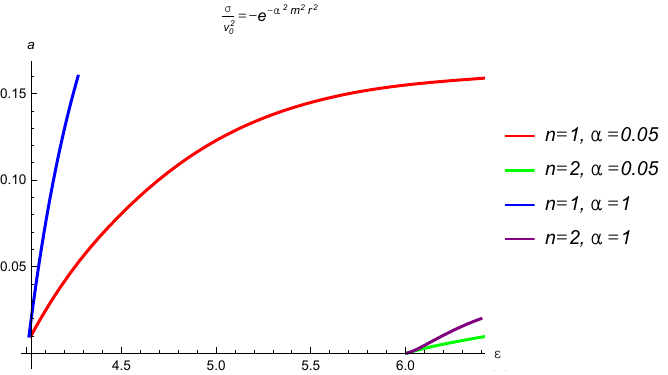}
\end{center}
\caption{Parameter $a$ in \eqref{054} as a function of 
the decay exponent $\varepsilon$ for a Gaussian impurity 
\eqref{sigmax} with different $n$ and $\alpha$. The lower
limits of $\varepsilon$ are almost $n+2$ which is the Liouville limit in the 
homogeneous theory.}
\label{fig:eab2}
\end{figure}

Numerically we find that if $\beta>0$ the lower bound of $\varepsilon$ does
not change much from $n+2$ which is the value obtained in Liouville 
approximation, as in the case of nontopological solitons discussed above. 
We plot $\varepsilon$ vs $a$ for $n=1,2$ in Figure~\ref{fig:eab2}.

\begin{figure}[H]
\begin{center}
        \includegraphics[page=3, width=0.55\textwidth]{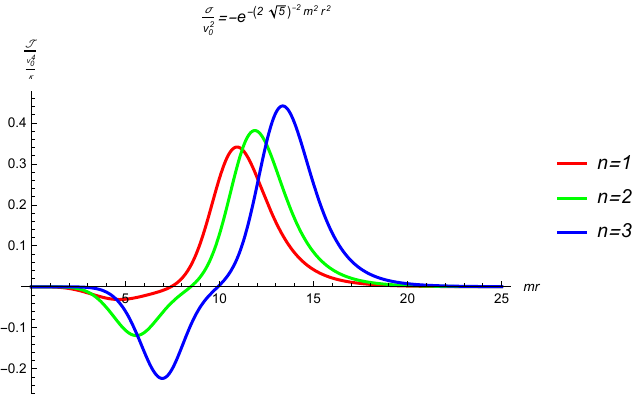}
\end{center}
\caption{Angular momentum densities of nontopological vortices in unit of 
$\kappa m^2$ for a Gaussian impurity \eqref{sigmax} with 
$\alpha^{-2}=20$ and $\beta=1$.}
\label{fig:nontopvorj}
\end{figure}

Nontopological vortices are spinning with angular momentum \eqref{phij}
which manifestly shows that it consists of two parts, namely the vortex 
contribution $-\pi\kappa^2 n^2$ and the Q-ball contribution 
$\pi \kappa \varepsilon^2$. This can also be seen from
the plot of the angular momentum density $\mathcal{J}$ in Figure 
\ref{fig:nontopvorj}. Note that the inner region (vortex part) and the outer 
region (Q-ball part) are spinning in different directions but the Q-ball
part mostly wins since $\varepsilon > n$.

\subsection{Solutions without rotational symmetry}

So far, we have considered solutions with rotational symmetry. Here we briefly
discuss general solutions which need not be rotationally symmetric. 
For topological vortices, the existence was shown in \cite{Han:2015tga} for
square integrable $\sigma(\boldsymbol{x})$. We counted the number of zero modes
with the standard procedure \cite{Jackiw:1990pr}
and confirmed that there are $2n$ zero modes
for topological vortices with vorticity $n$ as in the homogeneous case, which are identified with the $2n$ coordinates of $n$ vortex positions. 
We also checked that for nontopological solutions the number of zero modes
is $2(n+\hat\varepsilon)$ regardless of the impurity $\sigma$, where 
$\hat\varepsilon$ is the largest integer less than $\varepsilon$. 
Thus we expect that
there exist general multi-vortex solutions with arbitrary vortex points,
although index analysis does not completely prove the existence, that awaits further mathematical analysis.

\section{Conclusions}

In this paper, we have studied the CSH model with a magnetic
impurity which allows supersymmetric extension with a suitable choice of
the impurity term. Compared to the homogeneous theory without an impurity,
the number of supersymmetries is reduced from $\mathcal{N}=2$ to 
$\mathcal{N}=1$. Since the translation symmetry is broken, vacuum solutions
need not be constant in this theory. We showed that, in addition to the 
trivial vacuum $\phi=0$ in the symmetric phase, the vacuum in the broken 
phase has a nontrivial profile satisfying the Bogomolny equations. It has 
nontrivial energy density, magnetic field and angular momentum density
while its energy, magnetic flux and angular momentum remain all zero.
Depending on the sign of the impurity term added to the Lagrangian,
the Bogomolny equations only have vortex solutions with either positive 
vorticities or negative vorticities, but not both. 

As an explicit example,
we numerically obtained various type of rotationally symmetric soliton 
solutions such as (non)topological vortices and Q-balls for a Gaussian 
impurity. Basic properties of the solutions are similar to those in the
homogeneous theory and largely independent of the detailed form of the 
impurity, although there are some peculiarities in nontopological solitons
which are not protected by topology as discussed in section 4.
It is probably due to the fact that the impurity considered here is 
local and does not change the essential vacuum structure. It would be
interesting to see what would happen if one introduces inhomogeneities which
decay more slowly or do not vanish asymptotically. In this regards, we 
recall that in (1+1)-dimensional supersymmetric inhomogeneous theories
there is a rich spectrum of static BPS solutions~\cite{Kwon:2021flc}.

We also considered a delta-function impurity as a thin limit of
the Gaussian impurity. Thanks to the special feature of the BPS CSH model,
it does not alter the solutions except at the impurity point as far as 
the BPS solutions are concerned. It would be interesting to investigate 
further any implications of the result.

In this paper, we considered only magnetic impurities but we can
also include electric impurities as well as both electric and magnetic
impurities at the same time without breaking the BPS nature as discussed in
the Appendix A. It would then be a natural generalization to investigate
inhomogeneous models with these impurities. In this regards, we note that
CSH with constant background charge density was studied some 
time ago~\cite{Lee:1995eia} which may be considered as a constant electric 
impurity in this setup. 

In the homogeneous CSH model, there are also domain wall
solutions~\cite{Jackiw:1990pr} which connect the symmetric vacuum and
the broken vacuum. We expect that such solutions also exist in inhomogeneous
theories as long as the vacuum structure does not change. 
In this paper, we investigated theories with a U(1) Chern-Simons gauge field.
One can consider, instead, theories with a Maxwell 
term~\cite{Hook:2013yda,Tong:2013iqa,Kim:2024gfn} or both. The gauge
field can also be nonabelian~\cite{Kim:2019kns}.
We will report the results on these issues in separate publications.
 
\section*{\bf Acknowledgement}

We would like to thank Seungjun Jeon for useful discussions
and J.~G.~F.~Campos for informing us of the paper \cite{Bazeia:2024fgo}.
This work was supported by the National Research Foundation
of Korea(NRF) grant with grant number NRF-2022R1F1A1074051 (C.K.), NRF-
2022R1F1A1073053 (Y.K.), RS-2019-NR040081 (Y.K., O.K.), and RS-2023-00249608 (O.K.).

\appendix

\section{\bf Off-Shell Formulation}
We discuss the inhomogeneous supersymmetric CSH model in the off-shell 
formulation~\cite{Ivanov:1991fn, Schwarz:2004yj}. 
The homogeneous $\mathcal{N}=2$ supersymmetric Lagrangian is given by
\begin{align}
\mathscr{L}_\textrm{off-shell} 
 =&\, \frac\kappa2 \epsilon^{\mu\nu\rho} A_{\mu} \partial_{\nu} A_{\rho}
 + i\frac{\kappa}2\bar\chi \chi - \kappa N D 
 - \overline{D_{\mu} \phi} D^{\mu} \phi + i\bar{\psi} \gamma^{\mu} D_\mu \psi
\nonumber \\
 & + (|\phi|^2 - v_0^2)D - N^2 |\phi|^2 + |F|^2
 - i (\bar\psi \chi \phi + \bar\chi \psi \bar\phi) - i N \bar\psi \psi.
\label{lagoffshell}
\end{align}
It can easily be shown that it reduces to the on-shell Lagrangian \eqref{008} 
after elimination of the auxiliary fields.

Motivated by the fact that an abelian gauge field gives rise to a global
U(1) current via $\tilde{j}_\mu = \frac12 \epsilon_{\mu\nu\rho} F^{\nu\rho},$
we may consider turning on an external vector field $\hat{A}_\mu$
which minimally couples to the current $\tilde{j}_\mu \hat A^\mu$. 
It then results in a BF interaction 
$\frac12 \epsilon^{\mu\nu\rho} \hat{A}_\mu F_{\nu\rho}$. 
After supersymmetrization, we obtain~\cite{Hook:2013yda}
\begin{align}
\mathscr{L}_\textrm{BF} &= \frac12 \epsilon^{\mu\nu\rho} \hat{A}_\mu F_{\nu\rho}
  + \hat D N + \hat N D + \textrm{fermions} \nonumber \\
 &= \hat A_0 F_{12} + \hat N D + \hat F_{12} A_0 + \hat D N + \textrm{fermions},
\label{bflag}
\end{align}
where $\hat N$ and $\hat D$ denote the neutral scalar and the auxiliary 
field of the background vector multiplet, respectively. This term is invariant
under the following supersymmetric variation, 
which is the off-shell version of \eqref{005},
\begin{align}
\delta A_{\mu} &= \frac12 (\bar\eta \gamma_\mu \chi + \bar\chi \gamma_\mu \eta),
                         \nonumber \\
\delta N &= \frac12 (\bar\eta \chi + \bar\chi \eta), \nonumber \\
\delta D &= -\frac{i}2 (\bar\eta \gamma^\mu \partial_\mu \chi
                    - \partial_\mu \bar\chi \gamma^\mu \eta),
\end{align}
provided that the parameter $\eta$ satisfies the condition 
$\gamma^1 \eta = -i s \gamma^2 \eta$, where $s=\pm1$, and
\begin{align}
\hat A_0 &= -s\hat N \equiv s \sigma,  \nonumber \\
\hat F_{12} &= -s\hat D \equiv s \sigma_e .
\end{align}
Then, \eqref{bflag} becomes
\begin{equation} \label{bflag2}
\mathscr{L}_\textrm{BF}
 = \sigma ( s B - D) + \sigma_e (s A_0 - N)  + \textrm{fermions}.
\end{equation}

In this Lagrangian, $\sigma D$ term plays the role of an inhomogeneous version
of the Fayet–Iliopoulos term and nonzero $\sigma$ precisely 
corresponds to the magnetic impurity considered in the main part of the paper. 
On the other hand, nonzero $\sigma_e$ gives an electric impurity. In particular,
if the impurity is point-like, e.g. $\sigma_e = \delta(\boldsymbol{x})$, 
it simply reduces to a familiar $1/2$ BPS Wilson line. Thus, the electric 
impurity term in \eqref{bflag2} is possibly considered as a superposition 
of those Wilson lines. Note that both the electric and the magnetic 
impurities can be turned on simultaneously, while maintaining the supersymmetry.

\section{Notes on Delta-Function Impurity}

In this appendix, we choose the following form of the regularized delta 
function at the origin as the magnetic impurity,
\begin{equation} \label{deltatheta}
\tilde\delta(r) = \frac1{\pi\epsilon^2}\theta(\epsilon - r),
\end{equation}
where $\epsilon$ is assumed to be very small.
For $r < \epsilon$, the vacuum equation \eqref{deltavaceq} becomes
\begin{equation}
\nabla^{2} \ln |\phi|^{2}= \frac{4}{\kappa^{2}}|\phi|^{2} ( |\phi|^2 - v_0^2)
+\frac{4\eta}{\epsilon^2 v_0^2} |\phi|^2. 
                \label{deltavaceq2}
\end{equation}
When $\epsilon$ is small, it is clear that the second term on the right 
hand side dominates and \eqref{deltavaceq2} reduces to the Liouville equation.
It has a rotationally symmetric solution which is regular and
nonvanishing at the origin,
\begin{equation} \label{solin}
|\phi|^2 = \frac{2c^2 \epsilon^2 v_0^2}{\eta (c^2 - r ^2)^2},
\end{equation}
where $c$ is a constant. For $r > \epsilon$, the impurity term vanishes and
we expect $|\phi| \simeq v_0$. In this region, the solution is given by
the asymptotic behavior \eqref{asymp}.
Connecting the two solutions \eqref{solin} 
and \eqref{asymp} by requiring the continuities of $|\phi|$ and 
$d|\phi|/dr$ at $r=\epsilon$, we can obtain the constants $b$ and $c$
in the $\epsilon \rightarrow 0$ limit,
\begin{equation}
b\simeq \frac1{2\ln\frac1{m\epsilon}},
\qquad
c \simeq \left( \frac8\eta \epsilon^2 \ln\frac1{m\epsilon} \right)^{1/4} .
\end{equation}
Then, in the $\epsilon \rightarrow 0$ limit, we have
\begin{align}
|\phi(0)|^2 &= \frac{2 \epsilon^2 v_0^2}{\eta c^2} \rightarrow 0, \nonumber \\
|\phi(r)|^2 & \rightarrow v_0^2 \qquad (r \neq 0).
\end{align}
Therefore, in the presence of the magnetic delta-function impurity which 
is defined as the limit of the regularized function given by 
\eqref{deltatheta}, the vacuum solution is the constant solution $|\phi|=v_0$ 
except at the impurity point where $\phi$ vanishes, which is different from
the result of section 3 where $|\phi|=v_0$ including at the impurity point.
This verifies that the value of the scalar field depends on the regularization,
although physical quantities should not be affected by such a difference
at the single point.


\begin{thebibliography}{99}

\bibitem{Bogomolny:1975de}
E.~B.~Bogomolny,
Sov. J. Nucl. Phys. \textbf{24}, 449 (1976)
PRINT-76-0543 (LANDAU-INST.).

\bibitem{Prasad:1975kr}
M.~K.~Prasad and C.~M.~Sommerfield,
Phys. Rev. Lett. \textbf{35}, 760-762 (1975)
doi:10.1103/PhysRevLett.35.760

\bibitem{DiVecchia:1977nxl}
P.~Di Vecchia and S.~Ferrara,
Nucl. Phys. B \textbf{130}, 93-104 (1977)
doi:10.1016/0550-3213(77)90394-7

\bibitem{Witten:1978mh}
E.~Witten and D.~I.~Olive,
Phys. Lett. B \textbf{78}, 97-101 (1978)
doi:10.1016/0370-2693(78)90357-X

\bibitem{Bak:2003jk}
D.~Bak, M.~Gutperle and S.~Hirano,
JHEP \textbf{05}, 072 (2003)
doi:10.1088/1126-6708/2003/05/072
[arXiv:hep-th/0304129 [hep-th]].

\bibitem{Clark:2005te}
A.~Clark and A.~Karch,
JHEP \textbf{10}, 094 (2005)
doi:10.1088/1126-6708/2005/10/094
[arXiv:hep-th/0506265 [hep-th]].

\bibitem{DHoker:2007zhm}
E.~D'Hoker, J.~Estes and M.~Gutperle,
JHEP \textbf{06}, 021 (2007)
doi:10.1088/1126-6708/2007/06/021
[arXiv:0705.0022 [hep-th]].

\bibitem{DHoker:2006qeo}
E.~D'Hoker, J.~Estes and M.~Gutperle,
Nucl. Phys. B \textbf{753}, 16-41 (2006)
doi:10.1016/j.nuclphysb.2006.07.001
[arXiv:hep-th/0603013 [hep-th]].

\bibitem{Kim:2008dj}
C.~Kim, E.~Koh and K.~M.~Lee,
JHEP \textbf{06}, 040 (2008)
doi:10.1088/1126-6708/2008/06/040
[arXiv:0802.2143 [hep-th]].

\bibitem{Kim:2009wv}
C.~Kim, E.~Koh and K.~M.~Lee,
Phys. Rev. D \textbf{79}, 126013 (2009)
doi:10.1103/PhysRevD.79.126013
[arXiv:0901.0506 [hep-th]].

\bibitem{Kim:2018qle}
K.~K.~Kim and O.~K.~Kwon,
JHEP \textbf{08}, 082 (2018)
doi:10.1007/JHEP08(2018)082
[arXiv:1806.06963 [hep-th]].

\bibitem{Kim:2019kns}
K.~K.~Kim, Y.~Kim, O.~K.~Kwon and C.~Kim,
JHEP \textbf{12}, 153 (2019)
doi:10.1007/JHEP12(2019)153
[arXiv:1910.05044 [hep-th]].

\bibitem{Arav:2020obl}
I.~Arav, K.~C.~M.~Cheung, J.~P.~Gauntlett, M.~M.~Roberts and C.~Rosen,
JHEP \textbf{11}, 156 (2020)
doi:10.1007/JHEP11(2020)156
[arXiv:2007.15095 [hep-th]].

\bibitem{Kim:2020jrs}
Y.~Kim, O.~K.~Kwon and D.~D.~Tolla,
JHEP \textbf{12}, 060 (2020)
doi:10.1007/JHEP12(2020)060
[arXiv:2008.00868 [hep-th]].

\bibitem{Hook:2013yda}
A.~Hook, S.~Kachru and G.~Torroba,
JHEP \textbf{11}, 004 (2013)
doi:10.1007/JHEP11(2013)004
[arXiv:1308.4416 [hep-th]].

\bibitem{Tong:2013iqa}
D.~Tong and K.~Wong,
JHEP \textbf{01}, 090 (2014)
doi:10.1007/JHEP01(2014)090
[arXiv:1309.2644 [hep-th]].

\bibitem{Adam:2019yst}
C.~Adam, J.~M.~Queiruga and A.~Wereszczynski,
JHEP \textbf{07}, 164 (2019)
doi:10.1007/JHEP07(2019)164
[arXiv:1901.04501 [hep-th]].

\bibitem{Kwon:2021flc}
O.~K.~Kwon, C.~Kim and Y.~Kim,
JHEP \textbf{01}, 140 (2022)
doi:10.1007/JHEP01(2022)140
[arXiv:2110.13393 [hep-th]].

\bibitem{Ho:2022omx}
J.~Ho, O.~K.~Kwon, S.~A.~Park and S.~H.~Yi,
JHEP \textbf{11}, 219 (2023)
doi:10.1007/JHEP11(2023)219
[arXiv:2211.05699 [hep-th]].

\bibitem{Hong:1990yh}
J.~Hong, Y.~Kim and P.~Y.~Pac,
Phys. Rev. Lett. \textbf{64}, 2230 (1990)
doi:10.1103/PhysRevLett.64.2230

\bibitem{Jackiw:1990aw}
R.~Jackiw and E.~J.~Weinberg,
Phys. Rev. Lett. \textbf{64}, 2234 (1990)
doi:10.1103/PhysRevLett.64.2234

\bibitem{Wilczek:1981du}
F.~Wilczek,
Phys. Rev. Lett. \textbf{48}, 1144-1146 (1982)
doi:10.1103/PhysRevLett.48.1144

\bibitem{Arovas:1985yb}
D.~P.~Arovas, J.~R.~Schrieffer, F.~Wilczek and A.~Zee,
Nucl. Phys. B \textbf{251}, 117-126 (1985)
doi:10.1016/0550-3213(85)90252-4

\bibitem{Tsui:1982yy}
D.~C.~Tsui, H.~L.~Stormer and A.~C.~Gossard,
Phys. Rev. Lett. \textbf{48}, 1559-1562 (1982)
doi:10.1103/PhysRevLett.48.1559

\bibitem{Chen:1989xs}
Y.~H.~Chen, F.~Wilczek, E.~Witten and B.~I.~Halperin,
Int. J. Mod. Phys. B \textbf{3}, 1001 (1989)
doi:10.1142/S0217979289000725

\bibitem{Lee:1990it}
C.~Lee, K.~Lee and E.~J.~Weinberg,
Phys. Lett. B \textbf{243}, 105-108 (1990)
doi:10.1016/0370-2693(90)90964-8

\bibitem{Jackiw:1990pr}
R.~Jackiw, K.~Lee and E.~J.~Weinberg,
Phys. Rev. D \textbf{42}, 3488-3499 (1990)
doi:10.1103/PhysRevD.42.3488

\bibitem{Han:2015tga}
X.~Han and Y.~Yang,
JHEP \textbf{02}, 046 (2016)
doi:10.1007/JHEP02(2016)046
[arXiv:1510.07077 [hep-th]].

\bibitem{Bazeia:2024fgo}
D.~Bazeia, J.~G.~F.~Campos and A.~Mohammadi,
JHEP \textbf{12}, 108 (2024)
doi:10.1007/JHEP12(2024)108
[arXiv:2404.11694 [hep-th]].

\bibitem{Kim:1993mh}
C.~Kim, C.~Lee, P.~Ko, B.~H.~Lee and H.~Min,
Phys. Rev. D \textbf{48}, 1821-1840 (1993)
doi:10.1103/PhysRevD.48.1821
[arXiv:hep-th/9303131 [hep-th]].

\bibitem{Jaffe:1980mj}
A.~M.~Jaffe and C.~H.~Taubes,
{\em Vortices and Monopoles: Structure of Static Gauge Theories,} Birkh\"auser (1980).

\bibitem{Kim:1992yz}
Y.~Kim and K.~M.~Lee,
Phys. Rev. D \textbf{49}, 2041-2054 (1994)
doi:10.1103/PhysRevD.49.2041
[arXiv:hep-th/9211035 [hep-th]].

\bibitem{Kim:2024gfn}
Y.~Kim, S.~Jeon, O.~K.~Kwon, H.~Song and C.~Kim,
[arXiv:2409.12451 [hep-th]].

\bibitem{Lee:1995eia}
K.~M.~Lee and P.~Yi,
Phys. Rev. D \textbf{52}, 2412-2421 (1995)
doi:10.1103/PhysRevD.52.2412
[arXiv:hep-th/9501043 [hep-th]].

\bibitem{Ivanov:1991fn}
E.~A.~Ivanov,
Phys. Lett. B \textbf{268}, 203-208 (1991)
doi:10.1016/0370-2693(91)90804-Y

\bibitem{Schwarz:2004yj}
J.~H.~Schwarz,
JHEP \textbf{11}, 078 (2004)
doi:10.1088/1126-6708/2004/11/078
[arXiv:hep-th/0411077 [hep-th]].


\end{thebibliography}
\end{document}